\documentclass[aps,prl,twocolumn,showpacs,superscriptaddress,floatfix]{revtex4-1}
\usepackage{graphicx,graphics}
\usepackage{dcolumn}
\usepackage{amsmath,amssymb,amsfonts}
\usepackage{latexsym,verbatim}
\usepackage{bm}
\usepackage{color}
\usepackage[breaklinks=true,colorlinks,citecolor=blue,linkcolor=blue,urlcolor=blue]{hyperref}
\usepackage{tabularx}

\def\be{\begin{eqnarray}}
\def\ee{\end{eqnarray}}

\begin{document}
\title{Self-induced glassiness and pattern formation in spin systems subject to long-range interactions}
\author{Alessandro Principi}
\email{aprincipi@science.ru.nl}
\affiliation{Institute for Molecules and Materials, Radboud University, Heijndaalseweg 135, 6525 AJ, Nijmegen, The Netherlands}
\author{Mikhail I. Katsnelson}
\affiliation{Institute for Molecules and Materials, Radboud University, Heijndaalseweg 135, 6525 AJ, Nijmegen, The Netherlands}
\begin{abstract}
We study the glass formation in two- and three-dimensional Ising and Heisenberg spin systems subject to competing interactions and uniaxial anisotropy with a mean-field approach. In three dimensions, for sufficiently strong anisotropy the systems always modulates in a striped phase. Below a critical strength of the anisotropy, a glassy phase exists in a finite range of temperature, and it becomes more stable as the system becomes more isotropic. In two dimension the criticality is always avoided and the glassy phase always exists.
\end{abstract}
%
\maketitle

{\it Introduction}---The problem of pattern formation appears ubiquitously in many different fields, spanning from micromagnetics, to high-$T_{\rm c}$ superconductivity, to biology and social sciences.~\cite{Cross_rmp_1993,Koch_rmp_1994,Seul_science_1995,Ng_prb_1995,Gollub_rmp_1999,Choe_prb_1999,Swindale_prsl_1982,Bogomolny_prl_2002,Brucas_prb_2004,Brucas_prb_2008,Hafermann_apl_2009} The systems may differ quite substantially from one another, but their macroscopic phenomenology looks very similar. Therefore, also the models used to describe them can be very similar.~\cite{Swindale_prsl_1982,Prudkovskii_epl_2006} For example, a stripe phase emerges when the Gaussian part of the free energy of soft modes has either isolated minima or a mexican-hat shape in reciprocal space. In the latter case, the pattern emerges as a result of the spontaneous breaking of the rotational symmetry.~\cite{Prudkovskii_epl_2006}

Even in the complete absence of disorder and in a certain range of the external parameters (temperature and magnetic field), the very same free-energy can give rise to a self-induced glass.~\cite{Monasson_prl_1995,Mezard_jpa_1996,Schmalian_prl_2000,Westfahl_prb_2001,Westfahl_prb_2003,Parisi_jcp_2005,Parisi_rmp_2010} This result breaks down the picture that minima in reciprocal space  necessarily imply a stripe order in real space. Patterns look completely chaotic, even though a careful analysis reveals a hidden structure. For example, in the case of the mexican-hat free energy, the glass emerges as a superposition of stripe patterns with fixed periodicity but arbitrary direction.~\cite{Schmalian_prl_2000,Principi_prb_2016_glass} The direction is not completely arbitrary when the free energy has only isolated minima, and the period is not fixed when the line of minima has a non-circular shape. The hidden structure is revealed by analyzing the structure factor, which exhibits sharp peaks in correspondence of the minima~\cite{Prudkovskii_epl_2006,Principi_prb_2016_glass}. 

It is important to understand which properties are relevant to the formation of both ordered and random patterns. Frustration, stemming from the impossibility to locally fulfill at the same time all constraints, plays a fundamental role~\cite{Mezard_Parisi_Virasoro} and can arise in several ways. Ising spins antiferromagnetically coupled and arranged in a triangular lattice are a classical example.~\cite{Wannier_pr_1950} It is impossible to minimize the energy by looking at each plaquette, because of the freedom to arrange one of the three spins without changing the total energy. This is a cooperative problem, and the minimum-energy configurations can be found only by considering the whole system at the same time.~\cite{Mezard_Parisi_Virasoro,Wannier_rmp_1945} The ground state turns out to be massively degenerate, because of the freedom to flip a fraction of the total number of spins (up to $1/3$ in the case of the triangular lattice) without changing the total energy.~\cite{Wannier_pr_1950} A similar situation is realized in $J_1-J_2$ antiferromagnetic models at special values of the ratio $J_1/J_2$.~\cite{Chandra_jpcm_1990,Oitmaa_prb_1996,Schmidt_jpcm_2007,Bishop_jpcm_2012,Starykh_repprogphys_2015,Kotov_philmag_2000,Sushkov_prb_2001}

Geometrically frustrated systems, although extremely interesting, exhibit a very complicated phenomenology.~\cite{Wannier_pr_1950,Chandra_jpcm_1990,Oitmaa_prb_1996,Schmidt_jpcm_2007,Bishop_jpcm_2012,Starykh_repprogphys_2015,Kotov_philmag_2000,Sushkov_prb_2001,Nussinov_prb_2004} Therefore, they do not offer the simple playground that allows to highlight the main features of pattern formation and self-induced glassiness. In this paper we focus instead on $D$-dimensional systems of spins arranged in non-frustrated lattices and subject to competing interactions.~\cite{Kittel_pr_1946,Garel_prb_1982,Yafet_prb_1988,MacIsaac_physica_1994,Ng_prb_1995,Choe_prb_1999,Prudkovskii_epl_2006,Principi_prb_2016_glass} Nearest-neighbor spins are ferromagnetically coupled and each spin interacts with the others by a long-range interaction.
The energy dispersion of these systems exhibits a peculiar $D-1$-dimensional surface of minima, which turns out to be important for the emergence of a glassy phase.~\cite{Principi_prb_2016_glass} The wavevectors that lie in the minimum-energy surface correspond to equivalent striped phases. The potential landscape in the space of configuration is extremely ``rough'' and exhibits an exponential number (in the number of lattice sites) of local minima.~\cite{Charbonneau_naturecomm_2014} The large number of metastable states compensates for their small statistical weight.~\cite{Mezard_Parisi_Virasoro} It has been shown that, for such systems, a glass phase emerges in a certain interval of temperatures or in-plane magnetic fields.~\cite{Principi_prb_2016_glass} Such phase is characterized by the presence of an anomalous Green's function (off-diagonal in replica space), and a finite configurational contribution to the entropy.

Anisotropies can lift the degeneracy of the surface of minima. This does not imply that the glass is necessary destroyed but, depending on the form of the anisotropy, the formation of ordered patterns may be favored and a phase transition occur below a certain temperature.~\cite{Schmalian_prl_2000}
To reduce the model to its minimum, we consider the following free-energy for $D=2,3$
\begin{eqnarray} \label{eq:H_momentum_space}
{\cal F} = \frac{1}{2}\sum_{{\bm q}} G_{0}^{-1}({\bm q}) {\bm s}_{{\bm q}} \cdot {\bm s}_{-{\bm q}} + i \sum_i \sigma_i (s_i^2 - 1)
~,
\end{eqnarray}
where ${\bm s}_{{\bm q}} = \sum_i e^{i{\bm q}\cdot{\bm r}_i} {\bm s}_i$, ${\bm s}_i$ is an $N_s$-component spin located at site $i = 1,\ldots, N_{\rm L}$, $\sigma_i$ is a slave field (Lagrange multiplier) that ensures that $s_i \equiv |{\bm s}_i| = 1$. Throughout this letter, energies are measured in units of $J q_0^{D-2}$. Here $J$ is the exchange parameter. Finally, $G_{0}^{-1}({\bm q}) = q_0^D \left(q^2/q_0^2 - 1\right)^2/4 + q_0^D \varepsilon^2_{0} \sin^2(\theta_{\bm q})$. The line of minima is located at $|{\bm q}|=q_0$, while the term proportional to $\varepsilon_0^2$ introduces an ``easy-axis'' anisotropy. $\theta_{\bm q}$ is angle formed with the ${\hat {\bm x}}$ (${\hat {\bm z}}$) axis for $D=2$ ($D=3$). When $\varepsilon_0\neq 0$ the degeneracy of the minima is lifted: the system prefers to order in a striped phase with momentum ${\bm q}_0 = \pm q_0{\hat {\bm x}}$ (${\bm q}_0 = \pm q_0{\hat {\bm z}}$) for $D=2$ ($D=3$). The energy dispersion around these isolated minima is quadratic, {\it i.e.} $ G_{0}^{-1}({\bm q}) \propto |{\bm q} - {\bm q}_0|^2$. This model, although very simplified, captures the physics of spin systems subject to competing interactions and easy-axis anisotropy relevant for the formation of ordered and chaotic patterns.

A couple of comments are now in order. First of all, one may argue that the shape of the line of minima, which in Eq.~(\ref{eq:H_momentum_space}) is chosen to be a circle, might be important and lead to some qualitatively different behavior. Within the mean-field theory used in this paper,~\cite{Georges_rmp_1996} the shape is an irrelevant detail and that the choice of a circle does not undermine the generality of our results. This is shown briefly in what follows. On the other hand, non-local corrections to the self-energy~\cite{Rubtsov_prb_2008,Rubtsov_annphys_2012} may in principle depend on the shape of the line of minima and and lead to new features, like e.g. the lifting of the degeneracy of the minima even in the absence of anisotropy. The study of these corrections is beyond the scope of this paper.

As we show in what follows, the quadratic energy dispersion is sufficient to introduce a patterned phase in the three-dimensional case, but not in two dimensions.
In the latter case a phase transition can occur only when the energy dispersion around the isolated minima goes as $|{\bm q}-{\bm q}_0|^\alpha$ with $\alpha<2$. In this case the same phenomenology found for the 3D case applies also in 2D. 

This paper is organized as follows. In the next section we give all the details of the analytical solution of the problem. We then show and discuss, in a separate section, the numerical solution of the equations, pointing out the differences between the two- and three-dimensional cases. Finally, we conclude the paper by summarizing the results and discussing future perspectives and applications of our work.


{\it Mean-field approach to self-induced glassiness}---The problem of self-induced glassiness has been the subject of a few works.~\cite{Schmalian_prl_2000,Westfahl_prb_2001,Westfahl_prb_2003,Principi_prb_2016_glass} We therefore discuss only briefly the general strategy, and we go straight to the heart of the problem at hand.
To study the self-induced glassiness, we introduce in the Hamiltonian~(\ref{eq:H_momentum_space}) the $N_s$-component symmetry-breaking field ${\bm \psi}({\bm r})$, which glues the spins to a given configuration.~\cite{Schmalian_prl_2000,Principi_prb_2016_glass} The strength of the coupling between spins and ${\bm \psi}({\bm r})$, $g$, tends to zero after the thermodynamic limit is taken. The resulting free energy ${\cal F}_{\bm \psi}$ is analogous to that of a spin system subject to an infinitesimal quenched disorder.~\cite{Parisi_jcp_2005} Introducing replicas, we average over the configurations of ${\bm \psi}({\bm r})$. In the spirit of self-induced glassiness, the free energy ${\cal F}_{\bm \psi}$ is averaged with a probability distribution induced by itself, {\it i.e.} $P \sim e^{-\beta {\cal F}_{\bm \psi}}$.~\cite{Schmalian_prl_2000,Principi_prb_2016_glass}
The averaged Hamiltonian has the form~(\ref{eq:H_momentum_space}) where the fields are now replicated, {\it i.e.} ${\bm s}_{\bm q}\to {\bm s}_{\bm q}^{\alpha}$ and $\sigma_i \to \sigma_i^\alpha$ ($\alpha=1,\ldots,N$ denotes replica indices), and the bare Green's function acquires infinitesimal off-diagonal elements in replica space ($\propto g$).

The slave field $\sigma_i$ introduces an interaction between spins at different wavevectors ${\bm q}$. It is precisely this interaction which induces {\it finite} off-diagonal components of the Green's function (in replica space), when the latter is calculated self-consistently.~\cite{Principi_prb_2016_glass} This is analogous to what happens, e.g., in the theory of superconductivity: finite off-diagonal components of the Green's function (in Nambu space) emerge when Eliashberg's equations are solved self-consistently.~\cite{Shrieffer_book}

Owing to the local form of the last term of Eq.~(\ref{eq:H_momentum_space}) and in a mean-field spirit, we assume the self-energy to be a local quantity and to have a simple form in replica space: ${\tilde \Sigma} = \Sigma_{K} \delta_{\alpha\beta} + \Sigma_{F}$. $\Sigma_G \equiv \Sigma_{K} + \Sigma_F$ ($\Sigma_F$) is the normal (anomalous) component of the self-energy in replica space. In turn, the full Green's function reads ${\tilde G}_{\alpha\beta}({\bm q}) = K({\bm q})\delta_{\alpha\beta} + F({\bm q})$, where $K({\bm q}) = \big[ G_{0}^{-1}({\bm q})+ q_0^D \Sigma_{K} \big]^{-1}$ and $N F({\bm q}) =\big[ G_{0}^{-1}({\bm q}) + q_0^D(\Sigma_{K} + N \Sigma_{F}) \big]^{-1} - K({\bm q})$.
The self-energy is calculated by mapping the full model into the local problem~\cite{Westfahl_prb_2003}
\begin{eqnarray} \label{eq:action_local_problem}
{\cal H}_{\rm loc} =  \frac{1}{2}\sum_{\alpha,\beta} \Delta_{\alpha\beta} {\bm s}_{\alpha} \cdot {\bm s}_{\beta} + i \sum_\alpha \sigma_\alpha (s_\alpha^2 - 1)
~,
\end{eqnarray}
where $\Delta_{\alpha\beta} = \Delta_{K} \delta_{\alpha\beta} - \Delta_{F}$. The Green's functions of the local problem reads ${\bar G}_{\alpha\beta} = {\bar K}\delta_{\alpha\beta} + {\bar F}$, where ${\bar K} = \big[ \Delta_{K}+ \Sigma_{K} \big]^{-1}$ and $N {\bar F} = \big[ \Delta_{K} - N \Delta_{F} + \Sigma_{K} + N\Sigma_{F}\big]^{-1} - {\bar K}$.
These are related to the Green's function of the full model by the mean-field relation $\sum_{\bm q} {\tilde G}_{\alpha\beta}({\bm q}) = {\bar G}_{\alpha\beta}$. This equations determine $\Delta_K$ and $\Delta_F$ as a function of $\Sigma_K$ and $\Sigma_F$. It is clear that, since ${\bar G}_{\alpha\beta}$ is related to the integral of ${\tilde G}_{\alpha\beta}({\bm q})$ over all ${\bm q}$, the shape of the line of minima is not important (as long as it is smooth). 
In the limit $N\to 1$ the mean-field equations lead to $\Delta_{K} = - \Sigma_{K} + {\cal I}^{-1}(\Sigma_K)$ and $\Delta_{F} = \Delta_K +\Sigma_G - {\cal I}^{-1}(\Sigma_G)$, where
\begin{eqnarray} \label{eq:I_def}
{\cal I}(x) &=&
\left\{
\begin{array}{ll}
{\displaystyle \frac{1}{\pi} \int_0^{\pi/2} \frac{d\varphi}{\big[x+\varepsilon_0^2 \sin^2(\varphi)\big]^{1/2}} } & {\rm for}~ D=2
\vspace{0.3cm}\\
{\displaystyle \frac{1}{2\pi \varepsilon_0} \arctan\left(\frac{\varepsilon_0}{\sqrt{x}}\right) } & {\rm for}~ D=3
\end{array}
\right.
~.
\end{eqnarray}
Note that the integral on the first line can be expressed in terms of full elliptic integrals.

We now derive the self-consistent which describe the mean-field glass transition. The partition function of Hamiltonian~(\ref{eq:action_local_problem}) is rewritten as
\begin{eqnarray} \label{eq:Z_N_def}
Z(N) = 
\int_{0}^{\infty} d\lambda~ W_{N_s}(\lambda) \Omega^N(\lambda)
~,
\end{eqnarray}
where $W_{N_s}(\lambda) = \omega_{N_s} \lambda^{N_s-1} e^{-\lambda^2/(2\beta \Delta_F)}/(2\pi\beta\Delta_F)^{N_s/2}$, $\omega_{N_s} = 2\pi^{N_s/2}/\Gamma(N_s/2)$ is the solid angle in $N_s$ dimensions, $\Gamma(x)$ is the Euler gamma function, and
\begin{equation}
\Omega(\lambda) = 
\int d^{N_s} {\bm s} ~e^{-[ \beta \Delta_{K} s^2 + 2 \lambda s \cos(\theta)]/2} ~\delta(s^2 - 1)
~.
\end{equation}
Therefore, $\Omega(\lambda) = 2 e^{-\beta \Delta_K/2} \cosh(\lambda)$ for $N_s=1$, $\Omega(\lambda) = 2\pi e^{-\beta \Delta_K/2} I_0(\lambda)$ for $N_s=2$, and $\Omega(\lambda) = 4\pi e^{-\beta \Delta_K/2} \sinh(\lambda)/\lambda$ for $N_s=3$. Here $I_0(x)$ is the modified Bessel function of the first kind. In the limit $N\to 1$ we get $Z(1) = \omega_{N_s} e^{-\beta (\Delta_K-\Delta_F)/2}$.
From the equalities $N_s({\bar K} + {\bar F}) = -(2/N) \partial \ln Z(N)/\partial \Delta_K$ and $N_s({\bar K} + N {\bar F}) = (2/N) \partial \ln Z(N)/\partial \Delta_F$, in the limit $N\to 1$ we get the following self-consistent equations: 
\begin{subequations} \label{eq:SC_eq}
\begin{eqnarray} 
\label{eq:SC_eq_a}
&&{\bar K} + {\bar F}  = (N_s T)^{-1} 
~,
\\
\label{eq:SC_eq_b}
&&{\bar F} = 
\int_0^\infty d\lambda ~\frac{W_{N_s}(\lambda)}{N_s T Z(1)} \ln\Omega(\lambda) 
\left[\frac{\lambda}{\beta \Delta_F} \frac{\partial \Omega(\lambda)}{\partial \lambda} - \Omega(\lambda)\right]
\nonumber\\
&&
\equiv {\bar F} \big[{\cal J}(\Sigma_F) + 1\big]
~,
\end{eqnarray}
\end{subequations}
which allow us to determine the normal ($\Sigma_{G}$) and anomalous ($\Sigma_F$) components of the self-energy as a function of the temperature $T$ and anisotropy parameter $\varepsilon_0$. For future purposes, in Eq.~(\ref{eq:SC_eq_b}) we have introduced the function ${\cal J}(\Sigma_F)$, which is defined in terms of the integral on its first line. In solving these equations we have to require $\Delta_F>0$. 
The mean-field configurational entropy is determined from the free-energy ${\bar {\cal F}}(N) =  -(T/N) \ln Z(N)$ as ${\bar S}_{\rm c} = (1/T) \lim_{N\to 1} \partial {\cal F}(N)/\partial N$. It reads
\begin{eqnarray}
{\bar S}_{\rm c} &=&
\ln \big[e^{\Delta_K/2} Z(1)\big] - \frac{1}{Z(1)} \int_0^\infty d\lambda ~ W_{N_s}({\tilde \lambda})
\Omega(\lambda) 
\nonumber\\
&\times&
\left\{ \ln\big[\Omega(\lambda)\big] + \frac{\partial \ln W_{N_s}(\lambda)}{\partial \Delta_F} \frac{\partial \Delta_F}{\partial N}\Bigg|_{N\to 1} \right\}
~.
\end{eqnarray}
The derivative of $\Delta_F$ is found by considering the equality
\begin{eqnarray} \label{eq:F_derivative_N}
&&
\sum_{\bm q} F({\bm q}) = \frac{1}{N} \big[ {\cal I}(\Sigma_{K} + N\Sigma_{F}) - {\cal I}(\Sigma_{K}) \big]
\nonumber\\
&&
=
\frac{1}{N} \left[\frac{1}{\Delta_K-N \Delta_F + \Sigma_K+N\Sigma_F} - \frac{1}{\Delta_K + \Sigma_K} \right]
~.
\end{eqnarray}
Since the derivative is taken at fixed $F({\bm q})$, $\partial F({\bm q})/(\partial N) = 0$. 
Differentiating both lines of Eq.~(\ref{eq:F_derivative_N}) and setting them equal to zero we determine $\partial\Sigma_F/(\partial N)$ and $\partial \Delta_F/(\partial N)$.
The final expressions are quite cumbersome and will not be reported here. 

\begin{figure}[t]
\begin{center}
\begin{tabularx}{\columnwidth}{X X}
\multicolumn{2}{c}{ \includegraphics[width=0.99\columnwidth]{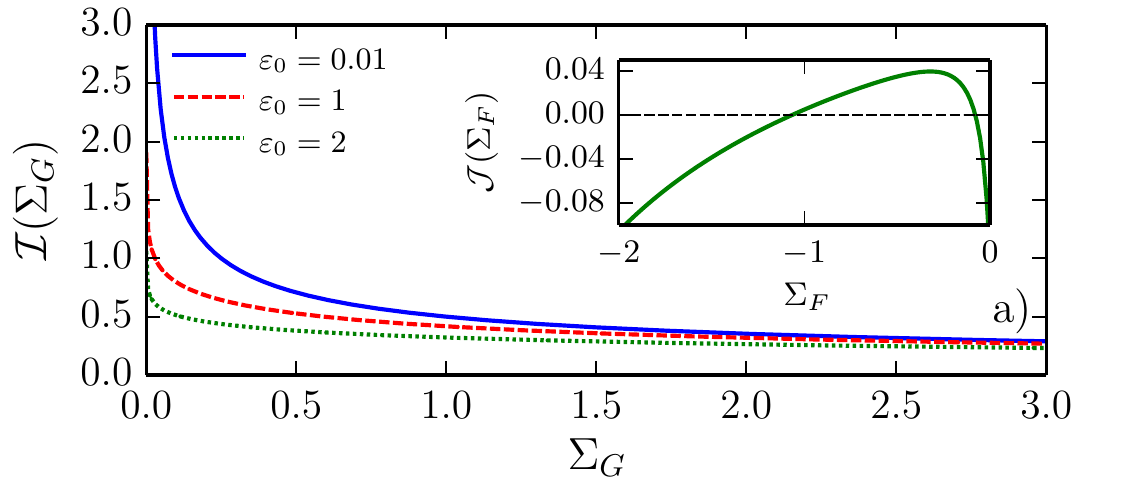} }
\\
\multicolumn{2}{c}{ \includegraphics[width=0.99\columnwidth]{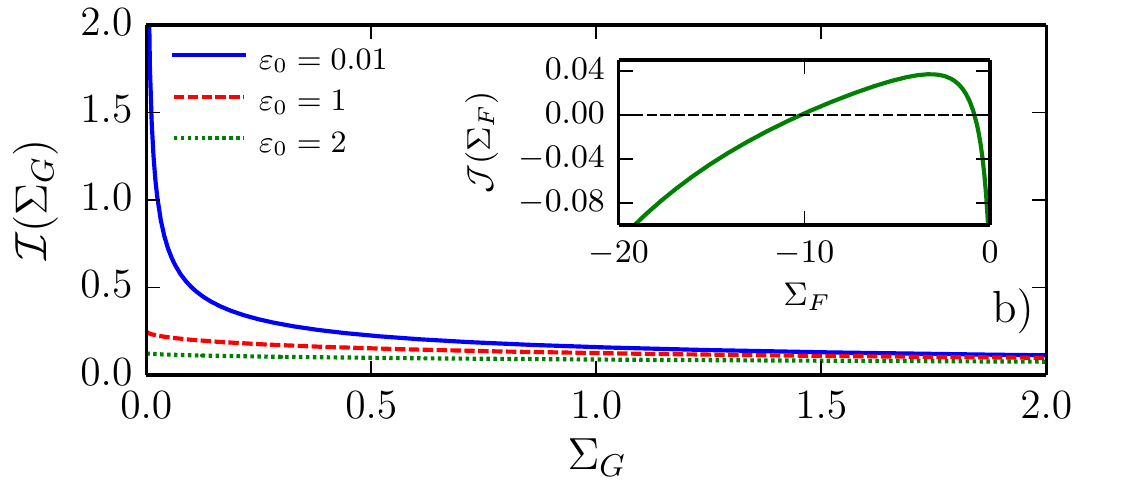} }
\end{tabularx}
\end{center}
\caption{
Panel a) the function ${\cal I}(\Sigma_G)$ for the two-dimensional Ising model ($D=2$ and $N_s = 1$), plotted as a function of $\Sigma_G$ and for three values of the anisotropy parameter $\varepsilon_0$. The function always diverges in the limit $\Sigma_G\to 0$. Inset: the function ${\cal J}(\Sigma_F)$ for $\varepsilon_0 = 0.01$ and $T=0.25$. It clearly shows two zeros.
Panel b) same as panel a) but for the three-dimensional Ising model ($D=3$, $N_s=1$). Note that, unless $\varepsilon_0=0$, the function always converges to a finite value, which defines the minimum temperature $T_{\rm crit}$ below which the system undergoes a phase transition to the ordered phase. Inset: the function ${\cal J}(\Sigma_F)$ for $\varepsilon_0 = 0.01$ and $T=2.5$.
\label{fig:one}}
\end{figure}

{\it Results}---We now consider Eqs.~(\ref{eq:SC_eq}) more closely. Using the definitions of $\Delta_K$ and $\Delta_F$ given after Eq.~(\ref{eq:I_def}), we rewrite Eq.~(\ref{eq:SC_eq_a}) as ${\cal I}(\Sigma_G) = (N_s T)^{-1}$. Since ${\cal I}(x)$ is a monotonous decreasing function, Eq.~(\ref{eq:SC_eq_a}) admits at most one solution for every temperature $T$ (at fixed $\varepsilon_0$). Therefore the value of $\Sigma_G$ is uniquely determined for any $T$ and $\varepsilon_0$. However, while in the two-dimensional case the function ${\cal I}(x)$ diverges for $x\to 0$, in three dimensions it reaches a finite value which scales as the inverse of the asymmetry parameter [compare Figs.~\ref{fig:one}a) and~b), main panels]. Therefore, in three dimension there exists a transition temperature $T_p$ such that Eq.~(\ref{eq:SC_eq_a}) can have a solution only for $T>T_p$. For $T<T_p$ the system orders in a striped phase. The stronger the asymmetry, the higher is $T_p$. Conversely, our model for $D=2$ has a transition temperature $T_p =0$.

The different behavior can be traced back to the fact that, for finite $\varepsilon_0$, the energy dispersion around the minima is quadratic and the mean-field equations are obtained by integrating the Green's functions ${\tilde G}_{\alpha\beta}({\bm q})$ over all momenta. Since a $1/q^2$-divergence is integrable in 3D but not in 2D, ${\cal I}(x\to 0)$ converges to a finite value in three dimensions and diverges when $D=2$. This is a situation of ``avoided criticality''.~\cite{Tarjus_jpcm_2005,Schmalian_prl_2000,Haule_prb_2007} The divergence is restored only in the isotropic case ($\varepsilon_0=0$), when the minima have an infinitely soft direction. Conversely, in 2D a phase transition to an ordered phase occurs if the energy dispersion goes as $\sim q^\alpha$ with $\alpha<2$ around the minimum or if the dispersion is not isotropic (e.g. quadratic in one direction and linear in the other).

Eq.~(\ref{eq:SC_eq_b}) defines also a temperature $T_A$ above which only the liquid phase can exist. 
We find no qualitative differences between the Ising, xy and Heisenberg models. Below $T_A$, Eq.~(\ref{eq:SC_eq_b}) admits two solutions for $\Sigma_F$. In the insets of Figs.~\ref{fig:one}a) and~b) we show the function ${\cal J}(\Sigma_F)$ for the two- and three-dimensional Ising models, respectively. Its zeros correspond to the values of $\Sigma_F$ which are solutions of Eq.~(\ref{eq:SC_eq_b}). Above $T_A$ no solution can be found and $\Sigma_F=0$.
 
\begin{figure}[t]
\begin{center}
\begin{tabularx}{\columnwidth}{X X}
\multicolumn{2}{c}{ \includegraphics[width=0.99\columnwidth]{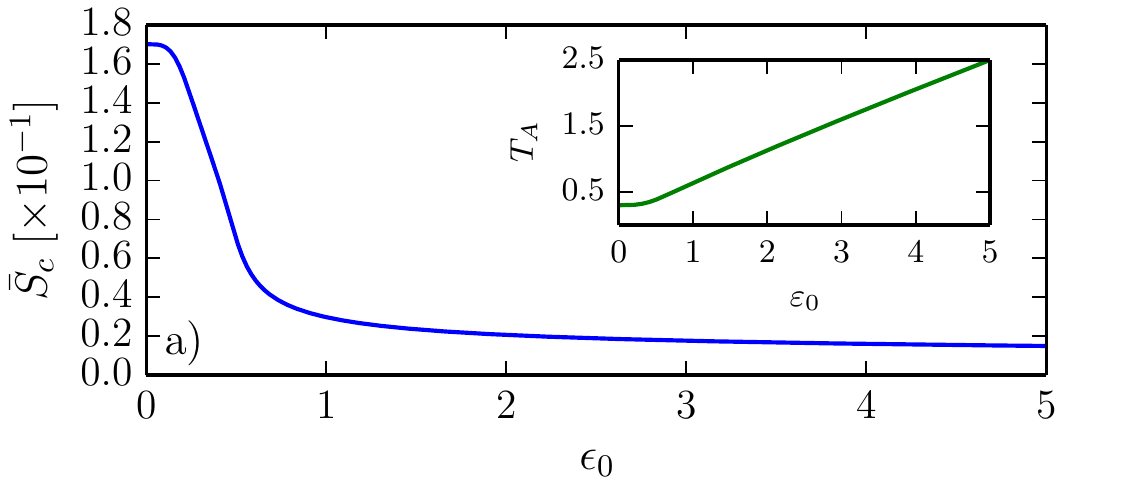} }
\\
\multicolumn{2}{c}{ \includegraphics[width=0.99\columnwidth]{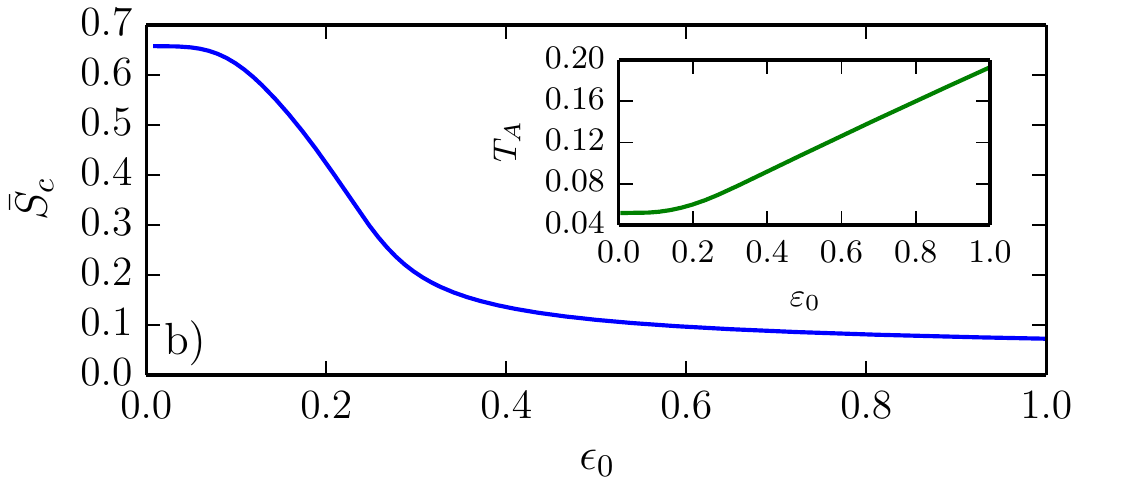} }
\end{tabularx}
\end{center}
\caption{
Panel a) the configurational entropy of the mean-field problem for the two-dimensional Ising model ($D=2$ and $N_s = 1$). Note that this curve has been multiplied by a factor $0.1$. Inset: the transition temperature $T_A$ as a function of the anisotropy parameter $\varepsilon_0$.
Panel b) same as panel a) but for the two-dimensional Heisenberg model ($D=2$, $N_s=3$). Inset: the temperature $T_A$ as a function of $\varepsilon_0$.
\label{fig:two}}
\end{figure}

In Fig.~\ref{fig:two} we address the stability of the two-dimensional Ising and Heisenberg glasses. From Eqs.~(\ref{eq:SC_eq_a})-(\ref{eq:SC_eq_b}) we calculate the liquid-glass transition temperature $T_A$ at which Eq.~(\ref{eq:SC_eq_b}) has only one solution. This is achieved by adding a third equation to the set, obtained by requiring the derivative of Eq.~(\ref{eq:SC_eq_b}) to vanish at $T_A$. The results for the two models are shown in the insets of Figs.~\ref{fig:two}a) and~b). Note that the transition temperature {\it increases} with the anisotropy parameter $\varepsilon_0$. 
As the isolated minima become deeper, higher temperatures are needed to introduce deformations and defects in the regular pattern.
Note also that the glass becomes more ``fragile'': the configurational entropy at the glass-liquid transition point (where it is maximum) decreases.

We quantify the fragility of the glass by calculating the configurational entropy of the mean-field problem ${\bar S}_c$. This is shown in the main panels of Fig.~\ref{fig:two}a)  and~b) for the two-dimensional Ising and Heisenberg models, respectively. The configurational contribution to the entropy decreases with increasing $\varepsilon_0$. This is expected, since the number of equivalent configurations should drastically decrease when the uniaxial anisotropy is introduced and the system is ``forced'' to assume a more ordered state. 

We stress again that the presence of a finite number of soft minima in momentum space is sufficient to avoid the ``critical behavior'' and the formation of an ordered phase in two dimensions. Therefore, at the mean-field level, the replica symmetry is always broken and a glass can always form. 
It is however well known that beyond-mean-field fluctuations can have a dramatic impact in two-dimensional systems. They can in principle destabilize the glassy phase and lead to transitions to other phases.
We expect the reduction in the value of the configurational entropy, already observed at the mean-field level, to become even more dramatic in the presence of fluctuations. A careful study of their role is beyond the scope of the present paper.

Finally, we note that the glassy phase is much more stable in the three-dimensional case, below the critical value of the anisotropy parameter $\varepsilon_0^{({\rm crit})}$. Indeed, the liquid-glass transition temperature and especially the configurational entropy remain nearly constant for $0<\varepsilon_0<\varepsilon_0^{({\rm crit})}$, {\it i.e.} $T_A = 2.96$ ($T_A = 0.51$) and ${\bar S}_c = 0.17$ (${\bar S}_c = 0.66$) for $N_s=1$ ($N_s=3$). Beyond the critical value $\varepsilon_0^{({\rm crit})}$ no glass can be realized and a transition to an ordered state always occurs, starting from the disordered (liquid) phase. 
We find that $\varepsilon_0^{({\rm crit})} = 0.33$ ($\varepsilon_0^{({\rm crit})} = 0.24$) for the three-dimensional Ising (Heisenberg) model.

{\it Summary and conclusions}---In this letter we studied the glass formation in two- and three-dimensional spin systems subject to competing short- and long-range interactions. In particular we analyzed within a mean-field framework the role of uniaxial anisotropy, which lifts the degeneracy of the line of minima in momentum space, leaving the system with few isolated ones. We find qualitative differences between the two- and three-dimensional cases. While in the former one criticality is avoided and a glass can always form, the latter undergoes a glass-ordered phase transition below a certain temperature. This result can be traced back to the softness of the energy dispersion around the minima. Moreover we find that, as the anisotropy is increased, the glass becomes more fragile and its configurational entropy decreases. Indeed, the number of equivalent configurations is expected to decrease (although it remains exponentially diverging in the mean-field limit and above the ordering temperature) and the energy landscape in the configuration space to smoothen.

The same phenomenology is expected to emerge in very different models, spanning from statistical physics, to information theory, biology and social sciences.~\cite{Cross_rmp_1993,Koch_rmp_1994,Seul_science_1995,Ng_prb_1995,Gollub_rmp_1999,Choe_prb_1999,Swindale_prsl_1982,Bogomolny_prl_2002,Brucas_prb_2004,Brucas_prb_2008,Hafermann_apl_2009,Prudkovskii_epl_2006,Monasson_prl_1995,Mezard_jpa_1996,Schmalian_prl_2000,Westfahl_prb_2001,Westfahl_prb_2003,Parisi_jcp_2005,Parisi_rmp_2010} In particular, we believe it to be relevant for the description of structural glasses of, e.g., hard spheres when the rotational symmetry in momentum space is broken.~\cite{Parisi_rmp_2010,Letz_pre_2000,DeMichele_prl_2007,Pfleiderer_epl_2008,Coluzzi_jcp_1999} 
Further investigation is needed to compare our results with those known in literature. It is however very promising that these models, in the isotropic case, exhibit a line of minima in momentum space,~\cite{Parisi_rmp_2010} and can therefore be regarded as ``stripe glasses''.

{\it Acknowledgments}---The authors acknowledge support from the ERC Advanced Grant 338957 FEMTO/NANO and from the NWO via the Spinoza Prize.

\bibliography{biblio}

\end{document}